\documentstyle[12pt]{article} 
\begin{document}
 \begin{center}
 {\bf Electromagnetic 2-forms on space-time} \\[2cm] 
M. Grigorescu \\[3cm]  
\end{center}
\noindent
$\underline{~~~~~~~~~~~~~~~~~~~~~~~~~~~~~~~~~~~~~~~~~~~~~~~~~~~~~~~~
~~~~~~~~~~~~~~~~~~~~~~~~~~~~~~~~~~~~~~~~~~~}$ \\[.3cm]
Two field 2-forms on the space-time manifold, in a relationship of duality, are presented
and included in the extended phase-space structure used to describe relativistic particles having both electric 
and magnetic charges. By exterior derivatives, these forms yield the two groups of Maxwell 
equations, while specific integrality conditions ensure magnetic monopole or electric charge 
quantization. Some properties of the common characteristic vector of the dual 2-forms are discussed. 
It is shown that the coupled energy-density continuity equation and the eikonal equation  
represent a classical, infinite-dimensional Hamiltonian system.
 \\
$\underline{~~~~~~~~~~~~~~~~~~~~~~~~~~~~~~~~~~~~~~~~~~~~~~~~~~~~~~~~
~~~~~~~~~~~~~~~~~~~~~~~~~~~~~~~~~~~~~~~~~~~}$ \\
{\bf PACS: 02.40.-k, 03.50.De, 42.15.-i   }  \\[1cm]

\newpage
\section{ Introduction} 
The geometric description of the electromagnetic field in terms of a connection form on the space-time manifold, 
related to the variation of length during parallel transport, was considered in \cite{weyl}. This connection form 
provides the Lorentz force, as it modifies the symplectic potential of the phase-space for an electric charge. Its 
curvature is a 2-form related to the first group of Maxwell equations, and Dirac's quantization condition for the 
magnetic monopole charge \cite{dirac, sni}. A second 2-form can also be defined by a duality relationship 
\cite{gs, bay}, and used to express the second group of Maxwell equations. 
\\ \indent
This work presents some properties of the 2-forms associated with the electromagnetic field, and their relevance 
for particle and photon dynamics. As these forms contain time as a coordinate rather than as a parameter, in 
Section 2 the motion of a relativistic electric charge is  described in terms of a Hamiltonian vector field on 
the cotangent bundle of the space-time manifold (the extended phase-space). The case of a particle having also 
a magnetic charge is considered in Appendix 1. The second 2-form and the Maxwell equations are presented in 
Section 3. The common characteristic vector of the 2-forms is studied in Section 4. It is shown that the 
photon dynamics in a transparent medium can be described as Hamiltonian flow of classical particles, with 
density and phase as canonically conjugate variables. Conclusions are summarized in Section 5.          
\section{Relativistic charge in extended phase-space} 
The canonical coordinates $q^e \equiv (q_0,{\bf q})$ and  $p^e \equiv (p_0,{\bf p})$ on the extended phase-space $M^e \equiv {\sf R}^8$ of a relativistic particle
consist of the canonical coordinates ${\bf q}=(q_1,q_2,q_3)$, ${\bf p}=(p_1,p_2,p_3)$ on the usual phase-space 
$M\equiv {\sf R}^6$, and $(q_0,p_0)$, supposed to be linear functions of time and energy, $q_0= c t$, 
respectively $p_0 = - {\cal E} /c$, where $c$ is a dimensional constant, identified with the speed of light in 
vacuum \cite{et}. As the convention of rising/lowering vector indices by the Lorentz metric tensor will not be 
applied, here and in the following all vector indices are subscripted, and all components are the physical ones. 
Let $u$ denote the "universal time" parameter along the trajectories on $M^e$, $d_u f \equiv d 
f/d u \equiv f^\prime$ the derivative of $f$ with respect to $u$, and $X_{H^e}$ the Hamiltonian vector field 
\begin{equation}
X_{H^e} = \sum_{\mu=0}^3 q_\mu^\prime \partial_\mu + p_\mu^\prime \partial_{p_\mu}~~,  \label{ld}
\end{equation}
$\partial_\mu  \equiv \partial / \partial q_\mu$,   
$\partial_{p_\mu} \equiv \partial / \partial p_\mu$, defined on $M^e$ by 
\begin{equation}
i_{X_{H^e}} \omega^e_0 =  d H^e~~, \label{hc}
\end{equation} 
with the canonical symplectic form $\omega^e_0 = - d \theta_0$,
\begin{equation}
\theta_0= \sum_{\mu =0}^3 p_\mu d q_\mu~~.
\end{equation} 

A free relativistic massive particle can be described using the extended 
Hamiltonian\footnote{
considered previously in \cite{sni}, p. 43, and independently in  \cite{rpw}.}
\begin{equation}
H^e_0= - c \sqrt{p_0^2 - {\bf p}^2}~~,
\end{equation}
so that (\ref{hc})  provides the equations of motion 
\begin{equation} 
q_0^\prime = - c \frac{p_0}{ \sqrt{p_0^2 - {\bf p}^2}}  ~~,~~ p_0^\prime = 0  \label{cle0}
\end{equation} 
\begin{equation} 
{\bf q}^\prime = c \frac{ {\bf p}}{ \sqrt{p_0^2 - {\bf p}^2}} ~~,~~ {\bf p}^\prime =0 ~~.
\label{cle1}
\end{equation} 
The usual veocity is ${\bf v} =  c {\bf q}^\prime/q_0^\prime = -c {\bf p}/p_0$, and the invariant value of $H^e_0=-m_0c^2$ defines the rest mass.   
\\ \indent
For an electric charge $e$, the coupling to the vector and scalar potentials of the electromagnetic field
 ${\bf A}=(A_1,A_2,A_3)$, respectively  ${\sf V}$, can be introduced by a local shift between mechanical and 
canonical momentum, so that $\omega^e_0$ becomes 
$$
\omega^e = - d \sum_{\mu =0}^3 (p_\mu + \frac{e}{c} A_\mu) d q_\mu ~~,
$$ 
where $A_0=-{\sf V}$. Thus $\omega^e= \omega_0^e + e \omega_f /c$ contains beside 
$\omega_0^e$ the field 2-form $\omega_f = - d \theta_f$,
\begin{equation}
\theta_f = \sum_{\mu=0}^3 A_\mu dq_\mu = {\bf A} \cdot d {\bf q} + A_0 d q_0~~.
\end{equation} 
The electric and magnetic fields ${\bf E} = - \partial_0 {\bf A} - \nabla {\sf V}$, and ${\bf B} = {\bf \nabla} \times {\bf A}$, therefore appear as coefficients of the 2-form $\omega_f$ on the space-time manifold ${\sf R}^4$ \cite{weyl}  
\begin{equation}
\omega_f = - \sum_{\mu < \nu} (\partial_\mu A_\nu - \partial_\nu A_\mu) d q_\mu \wedge dq_\nu \label{omf} 
\end{equation}
$$
= - {\bf B} \cdot d {\bf S} + {\bf E} \cdot d q_0 \wedge  d {\bf q}~~,
$$
where $dS_1 = dq_2 \wedge dq_3$, $dS_2 = - dq_1 \wedge dq_3$, $dS_3 = dq_1 \wedge dq_2$.  
The elements $[\omega_f]_{\mu \nu}=-( \partial_\mu A_\nu - \partial_\nu A_\mu)$ can also be represented in the matrix form
\begin{equation} 
[\omega_f] = \left[ \begin{array}{cccc} 
0 & E_1 & E_2 & E_3  \\ 
- E_1 & 0 & - B_3 & B_2 \\
- E_2 & B_3 & 0 & - B_1 \\
- E_3 & - B_2 & B_1 & 0 
\end{array} \right]~~. \label{m1}
\end{equation}
In the presence of the field, the equations of motion defined by $i_{X_{H^e}} \omega^e =  d H^e$ are 
\begin{equation} 
q_0^\prime = - c \frac{p_0}{ \sqrt{p_0^2 - {\bf p}^2}}  ~~,~~
p_0^\prime = - \frac{e}{c} {\bf E} \cdot {\bf q}^\prime  ~~,  
\end{equation} 
\begin{equation} 
{\bf q}^\prime = c \frac{ {\bf p}}{ \sqrt{p_0^2 - {\bf p}^2}} ~~,~~ 
{\bf p}^\prime = \frac{e}{c} ({\bf q }^\prime \times {\bf B} +  q_0^\prime {\bf E} ) ~~.
\end{equation} 
Denoting by $\dot{{\bf q}} \equiv {\bf v} =c {\bf q}^\prime/q_0^\prime$, $\dot{{\bf p}} \equiv c {\bf p}^\prime/q_0^\prime$, $\dot{{\cal E}} \equiv - c^2 p^\prime_0/q_0^\prime$ the usual derivatives of ${\bf q}$, ${\bf p}$, ${\cal E}$  with respect to the time $t = q_0/c$, these equations yield 
\begin{equation} 
\dot{{\bf p}} = \frac{e}{c} {\bf v } \times {\bf B} + e {\bf E}  ~~,~~
\dot{\cal E} =  e {\bf E} \cdot {\bf v}  ~~, \label{em1} 
\end{equation} 
with ${\bf v} = {\bf p} c^2 / {\cal E}$.
\section{The Maxwell equations} 
The field 2-form $\omega_f$ has an associated dual  $\omega_f^*$, which can be defined by
\begin{equation} 
\omega^*_f = \sum_{\mu < \nu, \alpha, \beta} (- \eta)^{\delta_{\alpha 0} + \delta_{\beta 0}} 
\epsilon_{\alpha \beta \mu \nu } \partial_\alpha A_\beta d q_\mu  \wedge dq_\nu  
= \eta {\bf E} \cdot d {\bf S} +  {\bf B} \cdot d q_0 \wedge d {\bf q}~~, \label{omfd}
\end{equation}
where\footnote{ $\sqrt{\eta}$ is the refractive index of the medium, presumed to be a positive constant. Though, metamaterials with negative refractive index have also been obtained \cite{pbs}.} $\eta = \epsilon_r \mu_r$,  $\epsilon_r$, $\mu_r$ denote the relative dielectric permittivity and magnetic permeability coefficients, $\delta_{\alpha \beta}$ is the Kronecker symbol, and $\epsilon_{\alpha \beta \mu \nu }$ is the unit tensor ($\epsilon_{0123 }=1$), fully antisymmetric to the permutations of the four indices. The elements $[\omega^*_f]_{\mu \nu}$ can be represented in matrix form as
\begin{equation}
[\omega^*_f] = \left[ \begin{array}{cccc} 
0 & B_1 & B_2 & B_3  \\ 
- B_1 & 0 & \eta E_3 & - \eta E_2 \\
- B_2 & - \eta E_3 & 0 & \eta E_1 \\
- B_3 & \eta E_2 & -  \eta E_1 & 0 
\end{array} \right]~~. \label{m2}
\end{equation} 
If $\lambda \in SO(1,3)^*$ is a Lorentz transformation, $\lambda^T \hat{g} \lambda = \hat{g}$,  $\hat{g} = diag [-1,1,1,1]$,
$$\tilde{q}_\mu = \sum_{\nu=0}^3 \lambda_{\mu \nu} q_\nu~~,~~\tilde{A}_\mu = \sum_{\nu=0}^3 (\lambda^{-1})^T_{\mu \nu} A_\nu~~,$$
then in normal vacuum\footnote{A non-trivial, subluminal refractive index in vacuum, might be induced by quantum 
gravitational fluctuations \cite{emn}.}  ($\eta=1$) 
\begin{equation} 
\sum_{\mu \nu \alpha \beta} (- 1)^{\delta_{\alpha 0} + \delta_{\beta 0}} 
\epsilon_{\alpha \beta \mu \nu } \tilde{\partial}_\alpha \tilde{A}_\beta d \tilde{q}_\mu  \wedge d \tilde{q}_\nu =  \sum_{\mu  \nu \alpha \beta} (- 1)^{\delta_{\alpha 0} + \delta_{\beta 0}} \epsilon_{\alpha \beta \mu \nu } 
\partial_\alpha A_\beta d q_\mu  \wedge dq_\nu ~~, \label{li1} 
\end{equation}
and $\omega^*_f$ is Lorentz-invariant (Appendix 3). 
\\ \indent
From (\ref{m1}) and (\ref{m2}) one obtains
\begin{equation}
{\rm det} [\omega_f] = \frac{1}{\eta} {\rm det} [\omega_f^*] = ({\bf E} \cdot {\bf B})^2 \label{det}
\end{equation}
while (\ref{omf}), (\ref{omfd}) yield 
\begin{equation}
\omega^*_f \wedge \omega_f = (\eta {\bf E}^2 - {\bf B}^2) d {\cal V}^e~~,
\end{equation}
$d {\cal V}^e = dq_0 \wedge d{\cal V}$, $d{\cal V} = dq_1 \wedge dq_2 \wedge dq_3$. \\ \indent
The first 2-form $\omega_f= - d \theta_f$ is exact, so that $d \omega_f =0$. This equality is equivalent to the first group of Maxwell equations \cite{weyl}
\begin{equation}
\nabla \cdot {\bf B} =0~~,~~{\bf \nabla} \times {\bf E} = - \frac{1}{c} \frac{ \partial{\bf B}}{\partial t} ~~.  \label{me1}
\end{equation}
It is important to remark that although true magnetic charges have not been observed, low-lying excitations 
resembling free magnetic monopoles can arise as defects in spin ice \cite{si1, si2, si3, si4}. The equations of 
motion for a quasiparticle which carries both electric and magnetic charges are derived in Appendix 1. In the 
presence of a magnetic charge $\omega_f$ is only locally exact by the Poincar\'e lemma \cite{am}, and the Dirac's 
quantization condition can be retrieved as an integrality condition for $e \omega_f /h c$ with respect to any 
space-like, compact, oriented, 2-dimensional surface (Appendix 2).  
\\ \indent  
If $\rho$ denotes the free electric charge density, integrable over ${\sf R}^3$, ${\bf j} = \rho \dot{{\bf q}}$, and  
\begin{equation}
J = \rho d {\cal V} - \frac{1}{c}  {\bf j} \cdot dq_0 \wedge  d {\bf S} \label{J}
\end{equation}
is an invariant 3-form, then the second group of Maxwell equations can be written 
as\footnote{A more general set of equations is provided by
$c d \omega_f^* = \mu_r i_{X_{H^e}} J^e $, with $J^e = \rho^e d{\cal V}^e$ and  
$\rho^e(q^e,u)$ the extended electric charge density, integrable over ${\sf R}^4$.}
\begin{equation}
d \omega_f^* = \mu_r J~~. \label{me2}
\end{equation}
Explicitly
\begin{equation}
d \omega_f^* = \eta \nabla \cdot {\bf E} d{\cal V} + ( \eta \partial_0 {\bf E} - {\bf \nabla} \times {\bf B} ) \cdot d q_0 \wedge d {\bf S} ~~,
\end{equation}
such that  (\ref{me2}) is equivalent to 
\begin{equation}
\epsilon_r \nabla \cdot {\bf E} = \rho ~~,~~
{\bf \nabla} \times {\bf B} = \frac{\mu_r}{c} ( {\bf j} + \epsilon_r \frac{\partial {\bf E}}{\partial t})~~.
\end{equation}
From (\ref{me2}) we also get $dJ =0$, which provides the continuity equation 
\begin{equation}
\partial_t \rho + {\rm div} {\bf j} =0~~,
\end{equation}   
and $ c \omega_f \wedge i_{\partial_0} d \omega_f^* = -\mu_r {\bf E} \cdot {\bf j} d q_0 \wedge d{\cal V} $, or explicitly 
\begin{equation}
\eta {\bf E} \cdot \partial_0  {\bf E} - {\bf E} \cdot {\bf \nabla} \times {\bf B} 
= - \frac{\mu_r}{c} {\bf E} \cdot {\bf j}~~.
\end{equation}
Replacing here ${\bf E} \cdot {\bf \nabla} \times {\bf B} = -{\rm div} ( {\bf E} \times {\bf B}) - {\bf B} \cdot \partial_0  {\bf B}$ from (\ref{me1}), 
we obtain
\begin{equation}
\partial_t w + {\rm div} {\bf Y}  +  {\bf E} \cdot {\bf j}=0~~, \label{wy}
\end{equation}
where $w = (\epsilon_r {\bf E}^2 + {\bf B}^2/ \mu_r)/2$ is the energy density of the field and ${\bf Y} = c {\bf E} \times {\bf B} / \mu_r$ the Poynting vector. \\ \indent
Worth noting is that when magnetic charges are included, the basic elements of the theory 
are the field 2-forms, rather than the local potentials $({\bf A},{\sf V})$, and the first 
group of Maxwell equations becomes $d \omega_f = - \mu_r J_m$, where $J_m$ is the 3-form 
(\ref{J}) for magnetic charges.  Also, in vacuum, an integrality condition for 
$\omega_f^*/ e$, where $e$ is a suitable constant, yields electric charge quantization 
(Appendix 2).  
 \section{The characteristic vector field and photon dynamics}
According to (\ref{det}), in general the rank of the electromagnetic 2-forms $\omega_f$, $\omega^*_f$ is not 
constant, and when ${\bf E} \cdot {\bf B} = 0$ both are degenerate. In the case ${\bf E} \cdot {\bf B} \ne 0$,  
$\omega_f$ is symplectic as it is closed by definition, while by (\ref{me2}), $\omega^*_f$ is closed only if $J=0$. 
\\ \indent 
Let us consider the common characteristic bundle $P_f$ over space-time,
\begin{equation}
P_f = \{ V \in T{\sf R}^4 / i_V \omega_f =i_V \omega_f^* =0 \} ~~. \label{cb} 
\end{equation}
Taking $V$ of the form $V = V_0 \partial_0 + {\bf V} \cdot \nabla$, $i_V \omega_f=0$ yields
\begin{equation}
{\bf E} \cdot {\bf V} =0~~,~~V_0 {\bf E} = - {\bf V} \times {\bf B} ~~, \label{c1}
\end{equation}
while from $i_V \omega_f^*=0$ one obtains
\begin{equation}
{\bf B} \cdot {\bf V} =0~~,~~V_0 {\bf B} = \eta {\bf V} \times {\bf E} ~~. \label{c2}
\end{equation} 
The equations (\ref{c1}), (\ref{c2}) hold also in inhomogeneous media, and
have a solution $V \ne 0$ only if ${\bf E} \perp {\bf B}$ and ${\bf B}^2= \eta {\bf E}^2 = \mu_r w$, when
\begin{equation}
{\bf V}^2 = \frac{1}{\eta} V_0^2 ~~,~~\frac{\bf V}{V_0}= \frac{{\bf Y}}{c w} ~~. \label{v1}
\end{equation}
Let us consider 
\begin{equation}
{\bf E} = k_0 {\bf P}_\varphi~~,~~{\bf B} = {\bf k} \times {\bf P}_\varphi~~, \label{eb}
\end{equation}
where ${\bf P}_\varphi$ provides the polarization,
\begin{equation}
k_0 = - \partial_0 \varphi ~~,~~{\bf k} = {\bf \nabla} \varphi~~,
\end{equation}
and $\varphi(q_0, {\bf q})$ is the phase function. In this case 
\begin{equation}
\omega_f = -  {\bf P}_\varphi \cdot d \varphi \wedge  d {\bf q}~~,
\end{equation} 
such that $i_V \omega_f =0$ if $i_V d \varphi =0$, or 
\begin{equation}
V_0 k_0 = {\bf V} \cdot {\bf k}~~. \label{v2}
\end{equation}
This equality, (\ref{v1}) and  
\begin{equation}
\frac{\bf k}{k_0}= \eta \frac{{\bf V}}{V_0}~~,
\end{equation}
obtained from (\ref{c2}), yield 
\begin{equation}
{\bf k}^2 = \eta k_0^2~~,
\end{equation}
or, in terms of $\varphi$, the eikonal equation
\begin{equation}
({\bf \nabla} \varphi)^2 = \eta (\partial_0 \varphi)^2~~. \label{eik}
\end{equation}
If $V \ne 0$ and ${\bf E}$, ${\bf B}$ are of the form (\ref{eb}), then in a non-dispersive,
transparent medium (${\bf j}=0$), the coupled equations (\ref{wy}), (\ref{eik}) ensure an extremum for
the action integral  
\begin{equation}
{\cal A}[n, \varphi] = - \int d^4 q ~ n [\partial_t \varphi +  \frac{c}{\sqrt{\eta}} \vert 
{\bf \nabla} \varphi \vert] \label{af} 
\end{equation}
with respect to the functional variations of the "photon density" $n(q_0, {\bf q})$ and $\varphi (q_0, {\bf q})$. Thus, $\delta_n {\cal A} =0$ yields (\ref{eik}) in the form 
\begin{equation}
\partial_t \varphi = -  \frac{c}{\sqrt{\eta}} \vert {\bf \nabla} \varphi \vert ~~, \label{eik1}
\end{equation}
while $\delta_\varphi {\cal A} =0$ provides  
\begin{equation}
\partial_t n + {\rm div} \lbrack n \frac{c}{\sqrt{\eta}} \frac{{\bf \nabla} \varphi }{
 \vert {\bf \nabla} \varphi \vert} \rbrack =0 ~~. \label{ce}
\end{equation}
In the stationary case $- \partial_t \varphi = ck_0  \equiv  \omega $ is a constant, 
$ \vert {\bf \nabla} \varphi \vert =\vert {\bf k}  \vert = \sqrt{\eta} k_0$, and (\ref{ce}) becomes
\begin{equation}
\partial_t n \omega + {\rm div} \lbrack n \frac{c^2}{\eta}{\bf \nabla} \varphi \rbrack=0~~.
\end{equation}
By multiplication with $\hbar$, considered as a dimensional factor converting  $\varphi$ into the "mechanical" action $S=\hbar \varphi$, this equation becomes (\ref{wy}) with \begin{equation}
w= n \hbar \omega ~~,~~ {\bf Y} = n \frac{c^2}{\eta} \hbar {\bf k} ~~,
\end{equation}
up to additive constants. Worth noting, (\ref{eik1}), (\ref{ce}) can also be expressed as an infinite-dimensional Hamiltonian system $i_{X_{\cal H}} \hat{ \omega} = {\sf d} {\cal H}~~$, where \cite{cpw}
\begin{equation}
\hat{ \omega}  = \int d{\cal V} {\sf d} n \wedge {\sf d} \varphi~~,~~ {\cal H}[n, \varphi]  = \int d{\cal V}  \frac{c}{\sqrt{\eta}} n \vert {\bf \nabla} \varphi \vert 
\end{equation}
and $n$, $\varphi$ are the conjugate variables. Single photons in inhomogeneous media 
therefore appear as classical particles with the canonical coordinates $({\bf q}, {\bf p}= \hbar {\bf k} )$, and the Hamilton function $h_{sp}({\bf q}, {\bf p}) = c \vert {\bf p} \vert / \sqrt{\eta({\bf q})}$. The equations describing their motion along the light rays take in this case the form considered before in \cite{mm}, 
\begin{equation}
\dot{\bf q} = \nabla_{\bf p} h_{sp} = \frac{c}{\sqrt{\eta}} \frac{{\bf p}}{\vert {\bf p} \vert}~~,~~ \dot{\bf p} = - \nabla_{\bf q} h_{sp} = h_{sp} \nabla \ln \sqrt{\eta} ~~.
\end{equation}  
The same equations can be obtained using (\ref{hc}), with an extended Hamiltonian $H^e_{sp} \equiv -c \sqrt{V_0^2 - \eta {\bf V}^2}= -c \sqrt{p_0^2 - {\bf p}^2/ \eta} $ defined in terms of the scalar $V_0 k_0 - {\bf V} \cdot {\bf k}$, normalized by $V_0 = \hbar k_0= - p_0$. 
\section{Summary and conclusions}
The geometric description of the elctromagnetic field using 2-forms on the space-time manifold arises in 
relativistic particles dynamics, or charge quantization. \\ \indent
In this work have been considered two field 2-forms, in a relationship of duality. The first 2-form 
$\omega_f = - d \theta_f$ is exact in the absence of the magnetic monopoles, and in Section 2 it was used to 
describe the motion of an electric charge as a Hamiltonian flow on the extended phase-space. The dual form was 
defined in Section 3 in terms of the field components and the refractive index of the medium. It is shown that this 
form modifies the symplectic potential of the extended phase-space for a magnetic charge, providing the 
"electric Lorentz force" (Appendix 1). The exterior derivatives of the two forms yield the two groups of Maxwell 
equations, while charge quantization can be introduced using specific integrality conditions (Appendix 2). In 
Section 4 the electromagnetic energy density and Poynting vector are related to the common characteristic vector 
($V$) of the dual 2-forms. By the dependence on the refractive index, this vector and the "wave-vector" ($k$), 
derived from the phase function, resemble the energy-momentum 4-vectors of Abraham, respectively of Minkowski. It 
is shown that the coupled energy-density continuity equation and the eikonal equation can be described as a 
classical, infinite-dimensional Hamiltonian system, with the photon density and the phase function as conjugate 
variables. Single photons appear as classical particles having as Hamiltonian a function $h_{sp}({\bf q}, 
{\bf p}) = \dot{\bf q} \cdot {\bf p}$, bilinear in velocity and momentum. Formally, one can also define an 
extended photon Hamiltonian, but further work is necessary to understand its significance, as in the limit of vanishing rest mass the universal time is not a suitable parameter.   
\section{Appendix 1: the equations of motion for electric/magnetic charges}
The equations of motion in vacuum for a relativistic massive particle which carries beside the electric charge 
$q_e$, a magnetic charge $q_m$, can be obtained replacing $\omega_0^e$ in (\ref{hc}) by
\begin{equation}
\omega^e = \omega_0^e + \frac{q_e}{c} \omega_f + \frac{q_m}{c} \omega_f^*~~. \label{ome}
\end{equation}      
In this case (\ref{em1}) are modified by the "electric Lorentz force" 
$- q_m{\bf v } \times {\bf E}/c $ and the magnetic field force $q_m {\bf B}$, such that 
one obtains
\begin{equation} 
\dot{{\bf p}} = \frac{q_e}{c} {\bf v } \times {\bf B} -  \frac{q_m}{c} {\bf v } \times {\bf E} + q_e {\bf E} + q_m {\bf B} ~~,~~
\dot{\cal E} =  (q_e {\bf E}+ q_m {\bf B}) \cdot {\bf v}  ~~. 
\end{equation} 
\section{Appendix 2: the charge quantization}
Let us consider the monopole vector field defined on ${\sf R}^3- \{ {\bf n}{\sf R}_- \}$,
$$ {\bf G}_n ({\bf r}) = \frac{a}{r} \frac{ {\bf n} \times {\bf e}_r }{1+{\bf n} \cdot {\bf e}_r}  $$
where $a$ is a constant, ${\bf n}$ is a fixed unit vector, and 
$${\bf e}_r \equiv  \frac{{\bf r}}{r}= ( \sin \theta \cos \varphi,  \sin \theta \sin \varphi ,  \cos \theta)~~,$$
with $r$, $\theta$, $\varphi$ the usual spherical coordinates of the position vector ${\bf r}$ in ${\sf R}^3$.
\\ \indent
Because $\nabla \times {\bf G}_{\bf n} = a {\bf e}_r / r^2$ independently of ${\bf n}$, a suitable set of local 1-forms $\alpha_{\bf n} = {\bf G}_{\bf n} \cdot d {\bf r}$
defines a symplectic form $\tilde{\omega}$ on the unit sphere ${\sf S}^2$, 
$$\tilde{\omega} \vert_{U_{\bf n}} \equiv d \alpha_{\bf n}  =( \nabla \times {\bf G}_{\bf n}) \cdot d {\bf S} = a \sin \theta d \theta \wedge d \varphi~~,$$
where the open set $U_{\bf n} = {\sf S}^2 - \{P_{-{\bf n}} \}$ is the domain of $\alpha_{\bf n}$, obtained by removing from ${\sf S}^2$ the "pole" $P_{-{\bf n}}$ located at ${\bf r}=-{\bf n}$.  Thus, if ${\bf n}$, ${\bf n}'$ are two distinct unit vectors, $\tilde{\omega} = d \alpha_{\bf n} =d \alpha_{{\bf n}'}$ on $U_{\bf n} \cap U_{{\bf n}'}$,  and the 1-form
$ \alpha_{\bf n} -  \alpha_{{\bf n}'} \equiv d \Phi_{{\bf n}{\bf n}'} $ is exact.  
For instance, if  ${\bf n} = {\bf k}$ and ${\bf n}' = - {\bf k}$, with ${\bf k} \equiv  (0,0,1)$, along the Z-axis, then
\begin{equation}
\alpha_{\bf k} =  a(1- \cos \theta) d \varphi~~,~~ 
\alpha_{-{\bf k}} = a(-1- \cos \theta) d \varphi
\end{equation}
and
\begin{equation}
 \alpha_{\bf k} - \alpha_{-{\bf k}} = 2 a d \varphi~~, \label{tf1}
\end{equation}
while by taking ${\bf n}' = {\bf i} \equiv (1,0,0)$, along the X-axis,
\begin{equation}
 \Phi_{{\bf k}{\bf i}} = a[ \varphi + \arctan ( \sin \varphi \tan \theta ) + \arctan ( \cot \varphi \cos \theta ) ]  ~~.
\end{equation}
In general, if $\alpha_i, \alpha_j$, defined on the open subsets $U_i, U_j \subset M$, are local 1-forms associated to a connection in a complex line bundle $L$ over the manifold $M$, with curvature $\tilde{\omega}$, then  on $U_i \cap U_j$ 
$$ \alpha_i - \alpha_j = \frac{1}{2 \pi i} \frac{d c_{ij}}{c_{ij}}~~,$$  
where $c_{ij} : U_i \cap U_j \mapsto {\sf C}$ are the transition functions \cite{bk}. For 
the unit circle subbundle of $L$, $c_{ij}$ become $c_{ij}^1 : U_i \cap U_j \mapsto {\sf U}(1)$ 
such that $c_{{\bf k} -{\bf k}}^1 = e^{ 4 \pi i a \varphi}$ provided by
(\ref{tf1}) is well-defined only if $4 \pi a =n$, $n \in {\sf Z}$.  In this case, if $\gamma \subset U_{\bf k} \cap U_{-{\bf k}}$ is any curve on ${\sf S}^2$, closed around the $Z$-axis, then
\begin{equation}
\int_{{\sf S}^2} \tilde{\omega} = \oint_\gamma (\alpha_{\bf k} - \alpha_{-{\bf k}})
 = 4 \pi a = n  ~~. \label{qc}
\end{equation} 
If the magnetic field in (\ref{ome}) is  ${\bf B} = \mu_r q_m' {\bf e}_r / 4 \pi r^2$, 
as expected for a point-like magnetic charge $q_m'$, and $\tilde{\omega} $ is identified with the space-like term $q_e {\bf B} \cdot d {\bf S} /hc$ of $\omega^e /h$, then in vacuum $a =  q_e q_m' / 4 \pi hc$ and (\ref{qc}) yields the Dirac's quantization condition 
\begin{equation}
q_e q_m'  = n hc~~,~~ n \in {\sf Z}~~. \label{dc}
\end{equation} 
One should note though that formally, the electric field ${\bf E} = q_e' {\bf e}_r / 4 \pi \epsilon_r r^2$ produced 
by a point-like electric charge $q_e'$ can also be expressed locally as ${\bf E}= q_e'\nabla \times {\bf G}_{\bf n} 
/ 4 \pi \epsilon_r a$. In such a case, in vacuum, an integrality condition for the term $q_m {\bf E} 
\cdot d {\bf S} /hc$ of $\omega^e /h$ reproduces (\ref{dc}) in the form $q_e' q_m  = n hc$, but if 
$\tilde{\omega} $ is identified with the space-like term in $\omega_f^* /e $ then $a= q_e'/ 4 \pi e$ and 
(\ref{qc}) provides electric charge quantization, $q_e' = ne$, $n \in {\sf Z}$.        
\section{Appendix 3: the Lorentz-invariance of $\omega_f^*$}
To prove (\ref{li1}),  let  $\lambda \in SO(1,3)^*$ be a Lorentz transformation, 
\begin{equation}
\lambda^T \hat{g} \lambda = \hat{g}~~,~~\hat{g} = diag [-1,1,1,1] \label{li2}~~,
\end{equation}
 such that
$$\tilde{q}_\mu = \sum_{\nu=0}^3 \lambda_{\mu \nu} q_\nu~~,~~\tilde{A}_\mu = 
\sum_{\nu=0}^3 (\lambda^{-1})^T_{\mu \nu} A_\nu~~.$$
If we denote $\tilde{\partial}_\mu \equiv \partial / \partial \tilde{q}_\mu$, then
$$\tilde{\partial}_\mu = \sum_{\nu =0}^3 (\lambda^{-1})^T_{\mu \nu} \partial_\nu~~,$$
and the left-hand side of (\ref{li1}) takes the form 
$$ 
\sum_{\alpha \beta \mu \nu} (- 1)^{\delta_{\alpha 0} + \delta_{\beta 0}} 
\epsilon_{\alpha \beta \mu \nu } \tilde{\partial}_\alpha \tilde{A}_\beta d \tilde{q}_\mu  
\wedge d \tilde{q}_\nu 
$$ 
$$
= \sum_{\alpha \beta \mu \nu} \sum_{ijkl} (- 1)^{\delta_{\alpha 0} + \delta_{\beta 0}} 
\epsilon_{\alpha \beta \mu \nu } (\lambda^{-1})^T_{\alpha i} (\lambda^{-1})^T_{\beta j}
\lambda_{\mu k} \lambda_{\nu l} \partial_i A_j dq_k \wedge dq_l~~.
$$
As $\hat{g}^2 =I$, from (\ref{li2}) we also get $\hat{g} (\lambda^{-1})^T = \lambda 
\hat{g}$, or in terms of the matrix elements 
$$(-1)^{\delta_{\alpha 0}} (\lambda^{-1})^T_{\alpha i} = \lambda_{\alpha i}  
(-1)^{\delta_{i 0}}~~,$$
and the previous expression becomes 
\begin{equation}
\sum_{\alpha \beta \mu \nu} \sum_{ijkl}  \epsilon_{\alpha \beta \mu \nu } 
\lambda_{\alpha i} \lambda_{\beta j} \lambda_{\mu k} \lambda_{\nu l} 
 (- 1)^{\delta_{i 0} + \delta_{j 0}} \partial_i A_j dq_k \wedge dq_l~~. \label{li3}
\end{equation}
However, $$ \sum_{\alpha \beta \mu \nu} \epsilon_{\alpha \beta \mu \nu }  \lambda_{\alpha i} 
\lambda_{\beta j} \lambda_{\mu k} \lambda_{\nu l} = \epsilon_{ijkl} {\rm det} ( \lambda) ~~,$$
whyle ${\rm det} ( \lambda ) =1$ because $\lambda \in SO(1,3)^*$, such that 
(\ref{li3}) becomes the right-hand side of (\ref{li1}).

\end{document}